# Kernel Adaptive Filtering for Nonlinearity-Tolerant Optical Direct Detection Systems

L. Zhang[(1,2)], O. Ozolins[(3)], R. Lin[(1)], A. Udalcovs[(3)], X. Pang[(1,3,6)], L. Gan[(4)], R. Schatz[(1)],
A. Djupsjöbacka[(3)], J. Mårtensson[(3)], U. Westergren[(1)], M. Tang[(4)], S. Fu[(4)], D. Liu[(4)], W. Tong[(5)],
S. Popov[(1)], G. Jacobsen[(3)], W. Hu[(2)], S. Xiao*[(2)], J. Chen*[(1)]

[(1)] KTH Royal Institute of Technology, Kista, Sweden, *jiajiac@kth.se*
[(2)] State Key Laboratory of Advanced Optical Communication System and Networks, Shanghai Jiao Tong University, Shanghai, China, *luzhang_sjtu@sjtu.edu.cn*, *slxiao@sjtu.edu.cn*
[(3)] NETLAB – Networking and Transmission Laboratory, RISE AB, Kista, Sweden
[(4)] Huazhong University of Science and Technology, Wuhan, China
[(5)] Yangtze Optical Fiber and Cable Joint Stock Limited Company, Wuhan, China
[(6)] Infinera, Fredsborgsgatan 24, 117 43 Stockholm, Sweden

**Abstract** *Kernel adaptive filtering (KAF) is proposed for nonlinearity-tolerant optical direct detection. For 7x128Gbit/s PAM4 transmission over 33.6km 7-core-fiber, KAF only needs 10 equalizer taps to reach KP4-FEC limit (BER@2.2e-4), whereas decision-feedback-equalizer needs 43 equalizer taps to reach HD-FEC limit (BER@3.8e-3).*

## Introduction

Nonlinearity is one of the key issues that hinders the development of high-capacity optical direct-detection (DD) systems[1]. The low-complexity linear channel equalization schemes, such as decision-feedback-equalizer with least-mean-square (DFE-LMS) algorithm, cannot perform properly to mitigate the system nonlinear distortions. Therefore, nonlinear equalization schemes (e.g. Volterra filtering[2-4], machine learning[5]) are needed. However, high computational complexity of these nonlinear equalization schemes is against the simplicity that is an inherent merit of optical DD systems.

Kernel method[6] is a new class of high dimension mapping schemes, where Mercer kernels can be utilized to produce high-dimension versions of the signals. Such a mapping makes a low-complexity linear adaptive filtering (LAF) mechanism possible for nonlinear equalization in optical DD systems.

In this paper, we propose a novel adaptive channel equalization scheme based on kernel method for nonlinearity-tolerant optical DD systems. By utilizing Mercer kernels, nonlinear noise in high-speed optical DD systems can be compensated by linear filtering mechanism, referred to as kernel adaptive filtering (KAF). In order to implement KAF, Kernel-LMS (KLMS) algorithm is introduced, which combines kernel method and LMS algorithm. Experimental demonstration shows KAF can significantly outperform conventional DFE-LMS while keeping low computational complexity. For 7x128Gbit/s pulse amplitude modulation with four amplitude levels (PAM4) transmission over 33.6km 7-core-fiber, employing KLMS with only 10 equalizer taps can achieve KP4-FEC[7] with bit error rate (BER) at 2.2e-4, whereas using DFE-LMS with 43 equalizer taps just reaches HD-FEC[7] with BER at 3.8e-3.

## Kernel adaptive filtering

The schematic diagram of KAF is shown in Fig. 1. KAF follows the classic sequential filtering for linear equalization, while using Mercer kernel as input signal mapping function.

Mercer kernel $\vartheta(\vec{c},\vec{c}\,')$ is a continuous and symmetric basis function defined in the kernel Hilbert space. In this paper, Gaussian kernel is utilized as the dominant expression of $\vartheta(\vec{c},\vec{c}\,')$:

$$\vartheta(\vec{c},\vec{c}\,') = e^{-\alpha\|\vec{c}-\vec{c}\,'\|^2}, \quad (1)$$

where $\vec{c}$ is the training data vector, $\vec{c}'$ is the measured data vector, and $\alpha$ is the Gaussian kernel bandwidth.

According to the Schölkopf representer theorem[8], the classic linear sequential processing has the universal approximation property for any continuous mapping function $f$ in kernel Hilbert space with Gaussian kernel expressed in Eq. 1. The corresponding mapping function $f$ can be expressed as follow:

$$f = \sum_{i=1}^{N} a_i \vartheta(\cdot, \vec{c}(i)), \quad (2)$$

where $N$ represents the number of training samples, and $a_i$ is the coefficient.

According to Eq. 2, it turns out that the mapping function can always be expressed in terms of the training data $\vec{c}$ with Mercer kernel. Therefore, the main idea of employing kernel method in nonlinear channel equalization is: 1) to transform the input data into a high-dimensional space by employing Eq.1, and 2) to apply an appropriate linear algorithm to process the inner product of the transformed input data and training data. Eq.1 can be expanded as follow[9]:

$$\vartheta(\vec{c},\vec{c}\,') = \sum_{i=1}^{\infty} \varepsilon_i \theta_i(\vec{c})\theta_i(\vec{c}\,'), \quad (3)$$

where $\varepsilon_i$ and $\theta_i$ are the non-negative eigenvalue and eigenfunction, respectively. A mapping $\varphi$ is denoted as a set of the eigenfunctions:

$$\varphi(\vec{c}) = \left[\sqrt{\varepsilon_1}\theta_1(\vec{c}), \sqrt{\varepsilon_2}\theta_2(\vec{c}),...\right]. \quad (4)$$

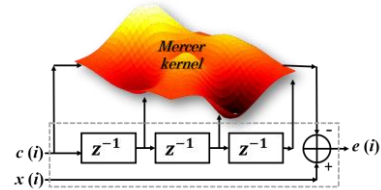

**Fig. 1:** Schematic diagram of KAF

$\varphi$ is the feature mapping and $\varphi(\vec{c})$ is the transformed feature vector in feature space. As a result, Eq. 1 can be expressed as follow:

$$\vartheta(\vec{c},\vec{c}\,') = \varphi(\vec{c})^T \varphi(\vec{c}\,'). \quad (5)$$

In KLMS, the training signal sequence $\vec{c}$ is transformed into $\varphi(\vec{c})$, which is then applied to the classic LMS mechanism. The $i$-th iteration of KLMS is expressed as:

$$\begin{cases} e(i) = x(i) - \vec{h}(i-1)^T \varphi(\vec{c}(i)) \\ \vec{h}(i) = \vec{h}(i-1) + \mu e(i)\varphi(\vec{c}(i)) \end{cases}, \quad (6)$$

where $x(i)$ is the desired training signal, $e(i)$ is the predicted error, $\vec{h}(i)$ is the estimated filter weight vector, and $\mu$ is the step-size parameter. By expanding the weight vector in Eq.6 iteratively, we can get:

$$\begin{aligned}\vec{h}(i) &= \vec{h}(i-1) + \mu e(i)\varphi(\vec{c}(i)) \\ &= \vec{h}(i-2) + \mu(e(i-1)\varphi(\vec{c}(i-1)) + \mu e(i)\varphi(\vec{c}(i))) \\ &\cdots \\ &= \vec{h}(0) + \mu\sum_{j=1}^{i} e(j)\varphi(\vec{c}(j)) \\ &= \mu\sum_{j=1}^{i} e(j)\varphi(\vec{c}(j)) \quad (\text{Assuming } \vec{h}(0)=0)\end{aligned} \quad (7)$$

Thus, after $i$-th step, the estimated filter weight vector is expressed as a linear combination of all the previous and present (transformed) inputs, multiplied by the predicted errors.

It is interesting to find that $\vec{h}(i)$ does not appear in the right side of Eq. 7. Instead, the sum of all past errors multiplied by the transformed feature vector of the previously received data (training data). Therefore, the equalization can be done by a single inner product, which saves a huge amount of computation time for nonlinear equalization in optical DD systems. Assuming $f_i$ is the equalization mapping at the $i$-th iteration, iterations in the KLMS algorithm can be expressed as follow:

$$f_{i-1}(\vec{c}(j)) = \mu\sum_{j=1}^{i-1} e(j)\vartheta(\vec{c}(j),\vec{c}(i)), \quad (8)$$

$$e(i) = x(i) - f_{i-1}(\vec{c}(i)), \quad (9)$$

$$f_i(\cdot) = f_{i-1}(\cdot) + \mu e(i)\vartheta(\vec{c}(i),\cdot), \quad (10)$$

where Eq. 8 is the *mapping function*, Eq. 9 is the *error function* and Eq. 10 is the *updating function*.

1. Set $\mu$ and choose kernel $\vartheta$, iteration number $N_L$;
2. for $i=1$ to $N_L$ do
   i. $f_{i-1}(\vec{c}(j)) = \mu\sum_{j=1}^{i-1} e(j)\vartheta(\vec{c}(j),\vec{c}(i))$,
   ii. $e(i) = x(i) - f_{i-1}(\vec{c}(i))$,
   iii. save $c(i)$ and $e(i)$,
3. end
4. **Nonlinear equalization output after $N_L$-th iteration:**
   $$f(\vec{c}\,') = \mu\sum_{j=1}^{N_L} e(j)\vartheta(\vec{c}(j),\vec{c}\,')$$

**Fig. 2:** KLMS algorithm for channel equalization

**Table. 1:** Complexity comparison at the $i$-th iteration

| Algorithm | Computation | Storage |
|---|---|---|
| LMS | O(L) | O(L) |
| KLMS | O(i) | O(i) |
| SVM | O($i^3$) | O($i^2$) |

The KLMS taps are indeed the number of dimensions of $\vec{c}$. The algorithm flow of KLMS for channel equalization is shown in Fig. 2.

The complexity comparison among different algorithms for channel equalization at the $i$-th iteration is shown in Table.1. Apart from LMS and KLMS, support vector machine (SVM)[5], as a classic machine learning algorithm, is also included for comparison. $L$ is the length of signals in LMS. The computation and storage complexity of KLMS are in the same order as that of LMS and perform obviously lower than that of SVM.

**Experimental setup and results**

The experimental setup for demonstrating the KLMS scheme is shown in Fig. 3. The 64Gbaud PAM4 signal is generated using 32Gbaud pulse pattern generator (PPG, Anritsu-MU183021A) and 64Gbaud digital-analog-converter (DAC, Anritsu-G0374A). A 1.55μm EML[10] with launch power of 1dBm is used to generate optical PAM4 signal. An Erbium doped fiber amplifier (EDFA) is used to amplify signal before a 1×8 splitter, decorrelation module and fan-in device. The signal transmitted over dispersion-compensated 33.6km single-mode 7-core fiber with low inter-core crosstalk (-45dB/100km). A fan-out device is used to couple the signals to standard single mode fiber. After the transmission, continuous fiber Bragg grating dispersion compensation module (DCM) with -672 ps/nm is used. A variable optical attenuator (VOA) is used to adjust the optical power after the pre-amplifier EDFA and before a PIN photodetector. An optical tunable filter (OTF) is used after pre-amplifier EDFA. PAM4 signal is captured with digital storage oscilloscope (DSO, 33GHz, 80GSa/s) for offline processing.

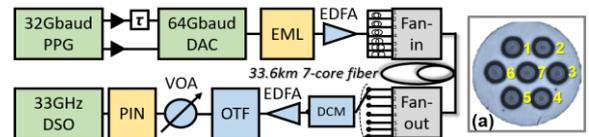

**Fig. 3:** Experimental setup, (a) the cross-view section of 33.6km 7-core fiber

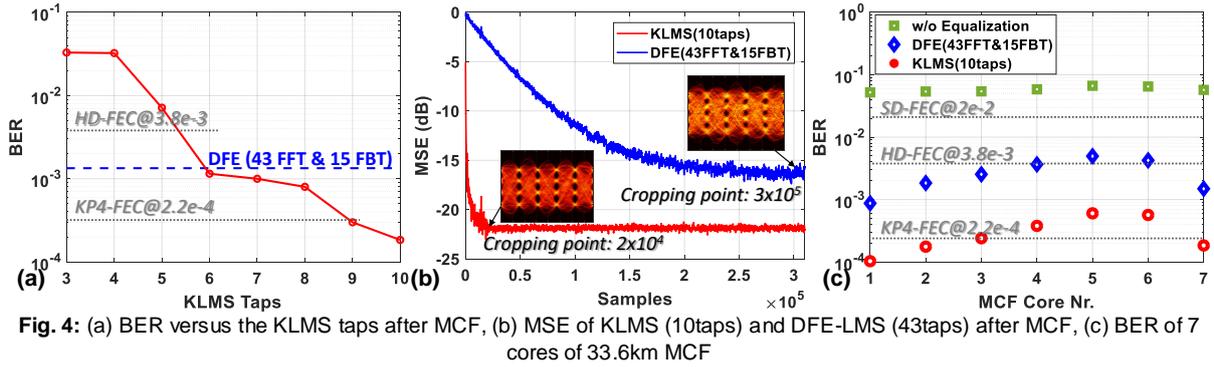

**Fig. 4:** (a) BER versus the KLMS taps after MCF, (b) MSE of KLMS (10taps) and DFE-LMS (43taps) after MCF, (c) BER of 7 cores of 33.6km MCF

The BER in terms of KLMS taps is shown in Fig. 4 (a). The BER is measured at the 7th-core after 33.6km transmission. Here, the number of cropping points (training samples) are set as $3 \times 10^5$. With the increase of KLMS taps, more nonlinear noise can be compensated and BER performance can be greatly improved. Compared with DFE-LMS that adopts 43 feed-forward taps (FFT) and 15 feedback taps (FBT), KLMS reaches the same performance with only 6 taps. With 10 taps, KLMS reaches KP4-FEC.

The training process in terms of mean-square-error (MSE) versus PAM samples is shown in Fig. 4 (b). Here, KLMS adopts 10 taps and DFE-LMS adopts 43 FFT and 15 FBT. The MSE is measured at the 7th-core. The cropping points of KLMS and DFE-LMS are set as $2 \times 10^4$ and $3 \times 10^5$ in experiment, respectively. It can be seen that KLMS outperforms DFE-LMS in terms of both convergence speed and MSE level. KLMS reaches -22dB MSE with ~ $2 \times 10^4$ samples and DFE-LMS reaches -17dB MSE with ~ $3 \times 10^5$ samples. Meanwhile, the taps of KLMS is much smaller than DFE-LMS, which can save more storage space and reduce the complexity. The recovered eye-diagrams are shown in the insets of Fig. 4 (b).

The BER performance of 7 cores after 33.6km transmission is shown in Fig. 4 (c). Without any equalization, the BER is higher than SD-FEC limit (i.e., 2e-2). After 43-taps DFE-LMS (43 FFT and 15 FBT), the BER of 7 cores are around HD-FEC limit (i.e., 3.8e-3). With KLMS, the BER of 7 cores is reduced to KP4-FEC limit (i.e., 2.2e-4). Here, the performance difference of 7 cores come from the insertion loss difference in the fan-in device.

## Conclusions

In this paper, we propose KAF for nonlinear equalization in optical DD systems. Experiments of 7x128Gbit/s PAM4 transmission over 33.6km 7-core-fiber demonstrates that KLMS reaches KP4-FEC (BER@2.2e-4) with only 10 taps while DFE-LMS reaches HD-FEC (BER@3.8e-3) with 43 taps. We believe KAF is a promising digital signal processing scheme for optical transmission, having a great potential to compensate nonlinear impairments not only in intensity modulated direct detection systems but also for coherent transmission, while keeping low complexity.

## Acknowledgements

We wish to thank the Swedish Research Council (VR), the Swedish Foundation for Strategic Research (SSF), Göran Gustafsson Foundation, the Swedish ICT-TNG, EU H2020 MCSA-IF Project NEWMAN (#752826), VINNOVA funded SENDATE-EXTEND and SENDATE-FICUS, National Natural Science Foundation of China (#61331010, 61722108, 61775137, 61671212).